# Novel Electrostatically Doped Planar Field-Effect Transistor for High Temperature Applications

Tillmann Krauss, Frank Wessely and Udo Schwalke

Institute for Semiconductor Technology and Nanoelectronics,
Technische Universität Darmstadt, 64289 Darmstadt, Germany

## Abstract

In this paper, we present experimental results and simulation data of an electrostatically doped and therefore voltage-programmable, planar, CMOS-compatible field-effect transistor (FET) structure. This planar device is based on our previously published Si-nanowire (SiNW) technology. Schottky barrier source/drain (S/D) contacts and a silicon-on-insulator (SOI) technology platform are the key features of this dual-gated but single channel universal FET. The combination of two electrically independent gates, one back-gate for S/D Schottky barrier modulation as well as channel formation to establish Schottky barrier FET (SBFET) operation and one front-gate forming a junctionless FET (JLFET) for actual current control, significantly increases the temperature robustness of the device.

## Introduction

The dominating leakage path for OFF-state currents in today's downscaled MOSFET devices originate from PN-junction and bulk leakage (1, 2). These leakage currents increase severely with temperature. SOI technologies can significantly reduce bulk leakage currents (13). However, junction leakage because of reversed bias PN-junctions is still present in SOI MOSFETs (2). Therefore, our device concept replaces conventional S/D PN-junctions with Schottky barrier contacts in a virtually dopant-free CMOS environment. Many groups have demonstrated the advantages of SBFET devices like low

S/D resistance as well as abrupt junctions enabling further scaling and suppression of parasitic bipolar behavior (19, 20). Furthermore, JLFET demonstrating high temperature robustness have been reported (16, 17, 18).

We have combined the advantages of the SBFET and JLFET concepts on SOI in a novel asymmetric dual gate but single channel device. This combination improves the high temperature robustness as reported recently for our SiNW FETs (3, 4). For the first time, we have successfully extended this 3D-SiNW concept to planar, non-nanowire device structures.

We found that the combination of an ambipolar electrostatic doping back-gate (BG) in addition to an electrically separated current flow control front-gate (FG) in a single device results in a superior on-to-off current ratio and leakage suppression at high temperatures. The BG transistor as a SBFET shows an ambipolar behavior similar to the SBFET reported in (21). On the other hand, the unipolar FG transistor resembles the behavior of a JLFET as reported in (18). Furthermore, as no conventional impurity doping process is required, the device does not suffer from dopant dependent reduction of carrier mobility or statistic dopant fluctuation and resulting threshold voltage variation following the argumentation of (16) for JLFET devices.

In addition, the degree of freedom to instantly select n- and p-type behavior via an ordinary electrical signal of appropriate polarity on the BG allows designing reconfigurable circuits with increased functionality (5). Until now, all ambipolar devices using a separate gate to control the current flow between source and drain involve nanowire structures as, for example, silicon nanowires or carbon nanotubes (6, 7).

To the best of our knowledge, this is the first proof of concept demonstrating a novel electrostatically doped, planar FET with an asymmetric dual-gate configuration which involves a combination of SBFET and JLFET properties for high temperature applications.

**Fabrication**

The field-effect transistors are fabricated on Premium Smart Cut™ SOI wafers from Soitec with a 70 nm top silicon and 145 nm buried oxide (BOX) layer doped with the lowest commercially available boron background doping of $10^{15}$ cm$^{-3}$ (8). A schematic cross section and an actually fabricated transistor are depicted in Fig. 1. Throughout this paper, device geometries are approximated by means of optical microscope only.

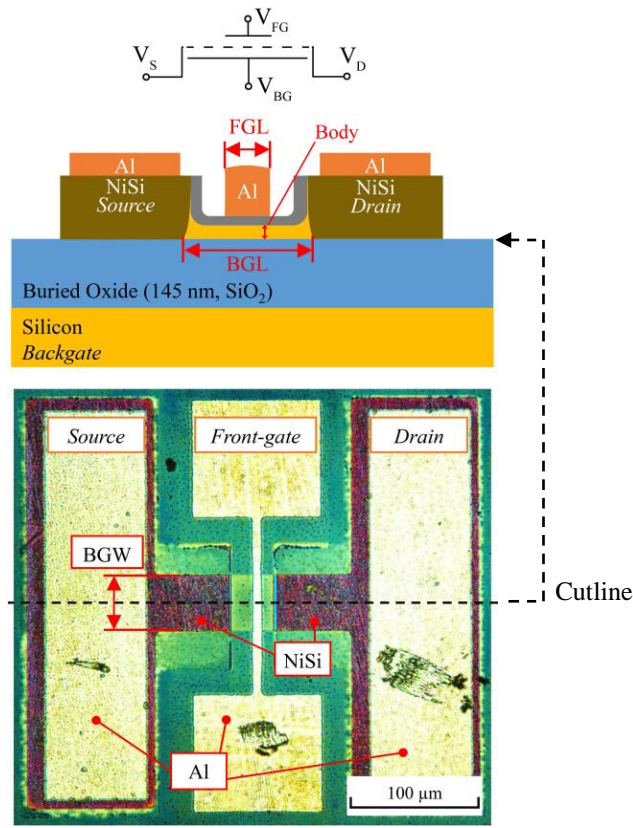

Fig. 1  Symbolic representation (top), schematic cross section of device structure (middle) as well as a top view optical micrograph of a fabricated device (bottom) with dimensions: back-gate length (BGL) 33.2 µm, front-gate length (FGL) 6.9 µm, back-gate width (BGW) 44.6 µm and body height approx. 20 nm.

The vertical structuring of the top silicon layer, including a FG recess formation, is conducted by Tetramethylammonium hydroxide (TMAH) wet etching (9). The time controlled TMAH recess etching results in a thinned body layer of approximately 20 nm within the FG region. The FG oxide of 7 nm is formed by dry thermal oxidation at 1000°C in a tube furnace followed by electron beam physical vapor deposition of 200 nm nickel. The S/D contacts are salicided in a horizontal tube furnace at 450°C for 15 minutes under forming gas ambient (90/10 $N_2/H_2$). During the forming gas process the metallic nickel diffuses into the silicon forming nickel-silicide (NiSi) near mid-gap Schottky-barrier junctions while the FG oxide serves as a diffusion barrier. Afterwards the excessive metallic nickel is stripped and an aluminum FG metal electrode and S/D contact pads are structured using a lift-off process. Note that during the entire fabrication no doping process is required (Fig. 2).

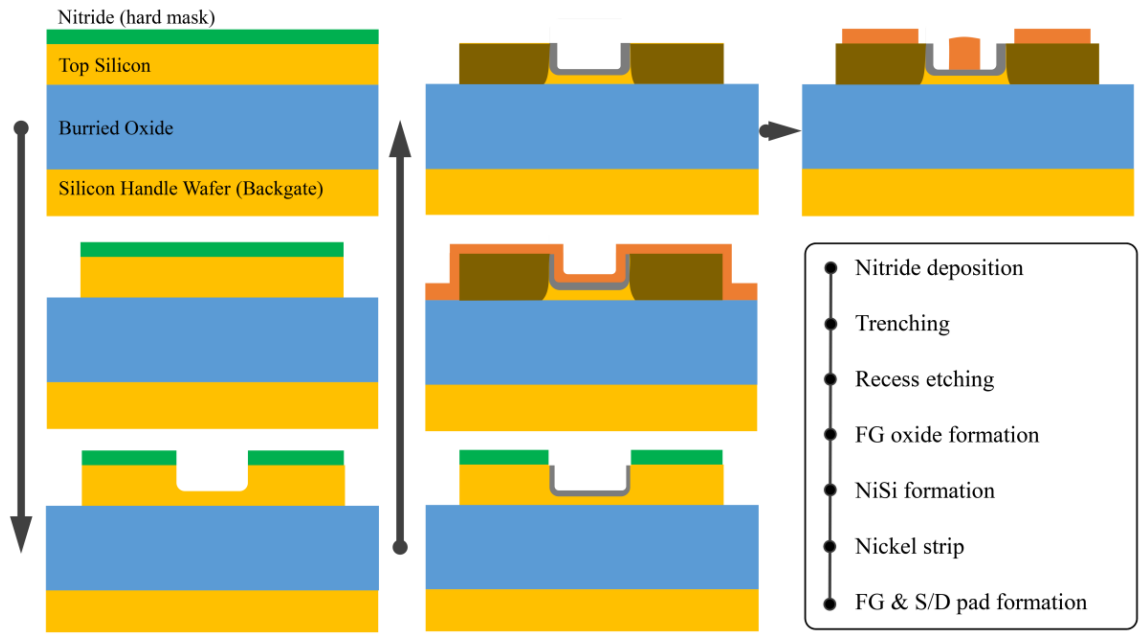

Fig. 2  Simplified fabrication process flow.

## Results and Discussion

To ease the understanding of the device operation, the device can be considered as an intrinsic composition of two entangled transistors, i.e. a JLFET on top of a SBFET. Each of these transistors is represented by its own gate electrode, the FG and BG respectively.

The BG electrode forms the first transistor and influences the whole channel region, i.e. body layer, as an ambipolar enhancement mode or normally-off SBFET (BGL in Fig. 1). The BG transistor defines the dominant charge carrier type, i.e. electrons or holes, via electrostatic doping. In contrast, the second transistor formed by the FG electrode influences only the center region of the channel (FGL in Fig. 1). The FG transistor resembles a depletion mode or normally-on JLFET and locally, i.e. in the area below the FG electrode, affects the formed channel of the BG transistor and therefore controls the charge carrier flow between source and drain. The FG is specially designed to avoid any parasitic FG channel formation by introducing a gap between FG and S/D hence preclude an overlap between S/D and FG that would normally be a necessity for conventional MOSFET structures. The combination of these two operation modes of enhancement and depletion can be appositely summarized as dehancement mode operation.

To obtain the intended operation of the dehancement mode, first an operating point has to be set by a DC voltage biasing of the BG electrode. This operating point determines the threshold voltage, maximum drain current and dominant charge carrier type of the dehancement mode transistor (DeFET). Afterwards, the actual switching behavior is realized by biasing the FG electrode similar to a depletion MOS transistor (Table 1).

| BG / FG Biasing | $V_{BG} \gg 0$ V – NMOS electron channel | $V_{BG} \ll 0$ V – PMOS hole channel |
|---|---|---|
| $V_{FG} \gg 0$ V | Conducting / on-state | Channel locally depleted / off-state |
| $V_{FG} \ll 0$ V | Channel locally depleted / off-state | Conducting / on-state |

Table 1  Dehancement mode operation states.

The supporting device simulations are based on standard drift-diffusion equations and universal Schottky barrier transport models excluding gate leakage models as reported in (22).

Back-Gate Enhancement Mode Operation

Fig. 3 depicts an experimental BG voltage sweep for a long channel device with an electrically floating FG and a constant drain/source bias of $V_{DS} = 1$ V displaying a clear ambipolar device characteristic. Applying a sufficiently high BG bias voltage ($V_{BG}$) results in an increasing drain current as either holes ($V_{BG} < -4$ V) or electrons ($V_{BG} > 13$ V) are attracted to the body-to-BOX interface forming a channel. The charge carrier concentrations are illustrated in Fig. 4 for positive and negative BG bias conditions. The simulated density of charge carriers lies within the range of $N = 10^{19}\,\text{cm}^{-3}$ in both cases (Fig. 4). As mentioned before, this resembles an ambipolar enhancement mode or accumulation mode transistor.

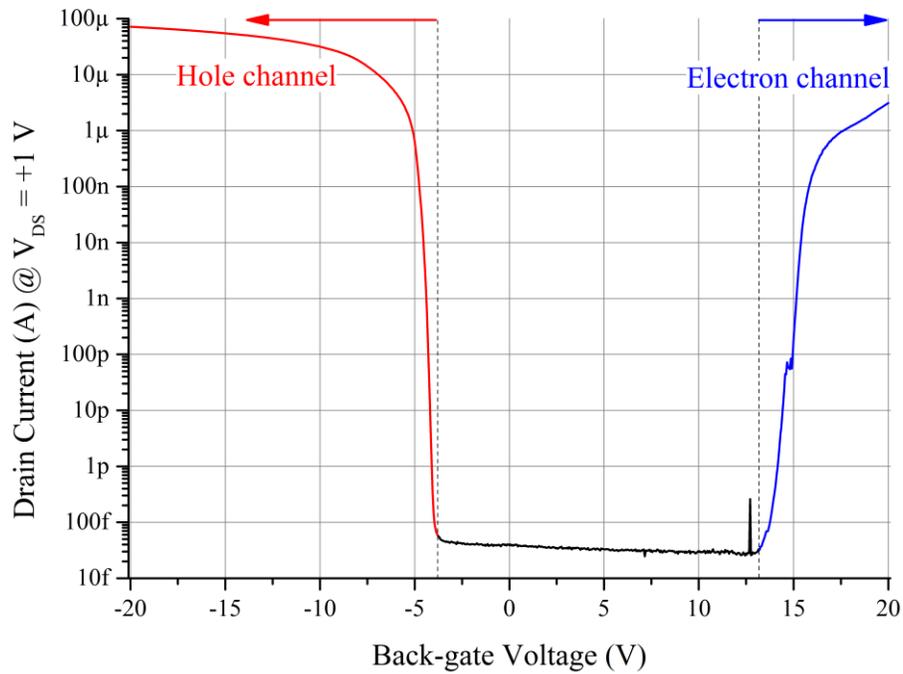

Fig. 3  Experimental back-gate voltage sweep, $V_{DS} = \pm 1$ V. Transistors dimensions: back-gate length 33.2 µm, back-gate width 44.6 µm.

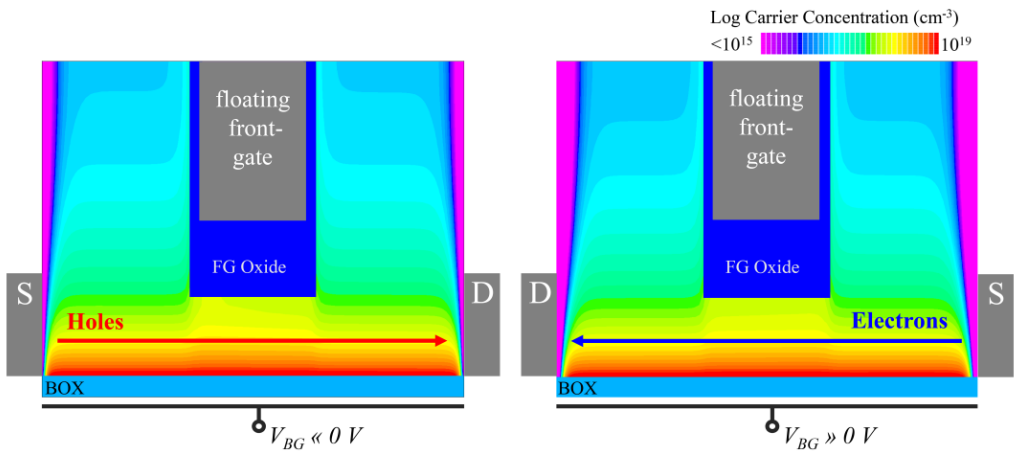

Fig. 4  Simulated charge carrier concentration in the top silicon layer with floating front-gate condition for electron ($V_{BG} \gg 0$ V) and hole ($V_{BG} \ll 0$ V) channel conduction.

The attracted charge carriers originate from the mid-gap Schottky barrier S/D contacts with an estimated work function of 4.6 eV. As typical for SBFET operation, the electrical field of the BG alters the shape especially width of the Schottky barriers and therefore influences the probability for tunneling and thermionic field emission of charge carriers through the Schottky barriers (19, 20). The parameters for the simulation of the Schottky barriers have been derived by fitting simulated and measured back-gate voltage sweeps of fabricated SiNW SBFET devices as described in (22). Deviations remain because of the complexity of the Schottky barrier interface especially the dopant profile due to dopant segregation (14, 15), geometrical differences and other not modeled effects, i.e. Fermi level pinning and interface states.

### Front-Gate Depletion Mode Operation

The measured input characteristics of the FG JLFET for various BG biasing conditions at room temperature are shown in Fig. 5 and Fig. 6. The input characteristics show a sub-threshold slope of 71 mV/dec (n-type) and 84 mV/dec (p-type) as well as an on-to-off drain current ratio of up to ~9 decades for the p-type and ~7 decades for the n-type mode FET. As expected, the maximum drain on-current is directly linked to the magnitude of back-gate voltage (e.g. compare on-currents in Fig. 5 and Fig. 6 with Fig. 3). The drain off-current on the other hand does not show any correlation to the applied BG potential stating a good local electrostatic control of the body region by the FG. A reduction of the body thickness from currently approximately 20 nm down to 5 nm is expected to exponentially decrease the off-current down to the single-digit fA range. Furthermore, the threshold voltage can be shifted by approx. 72 to 95 mV/$V_{BG}$ (Fig. 5). This threshold shift ability in conjunction with FG work function engineering for p- and n-type DeFET may result in additional design flexibility for circuit designers in low power applications (25).

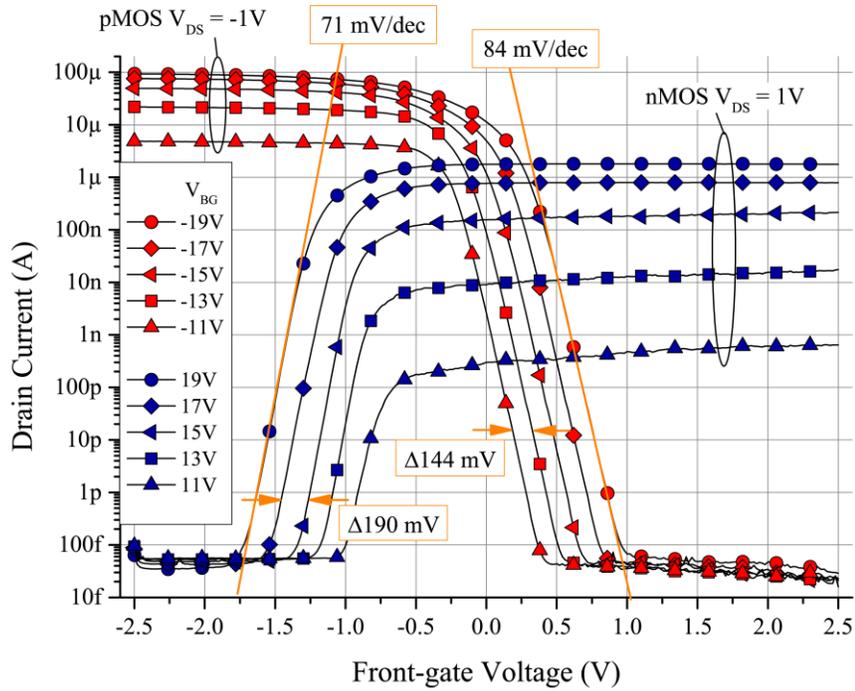

Fig. 5    Measured input characteristic of the fabricated front-gate transistor with an aluminum metal front-gate plotted on a logarithmic scale at room temperature. Transistor dimensions: BG length 33.2 μm, BG width 44.6 μm, FG length 6.9 μm and body thickness approx. 20 nm.

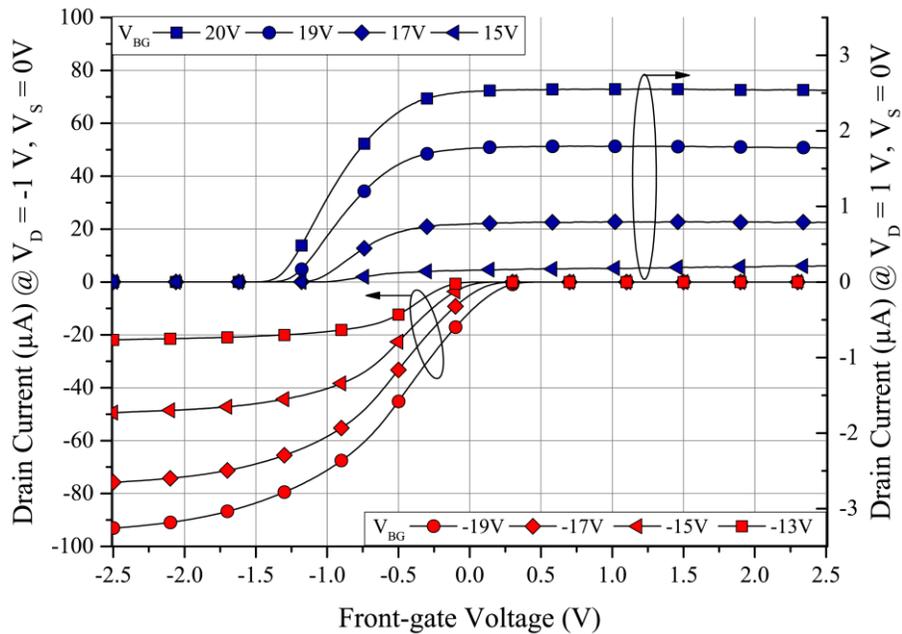

Fig. 6    Measured input characteristic of the fabricated front-gate transistor with an aluminum metal front-gate plotted on a linear scale at room temperature. Transistor dimensions: BG length 33.2 μm, BG width 44.6 μm and FG length 6.9 μm and body thickness approx. 20 nm.

The low leakage currents of less than 40 fA (~0.8 fA/µm) for p- and n-type operation are a result of high potential barriers for electrons and holes formed by the FG. According to simulations, these potential barriers reach up to 0.8 eV for silicon oxide as FG dielectric as illustrated in the simulated band diagram in Fig. 7 for n-type operation (vice versa for p-type operation – not shown here).

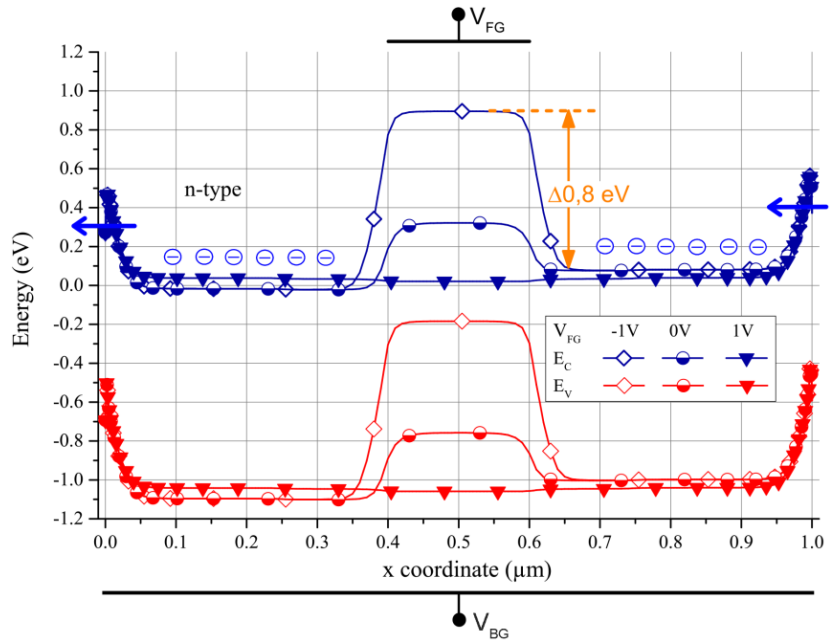

Fig. 7  Simulated band diagrams at the top silicon to buried interface for positive back-gate biasing $V_{BG} \gg 0$ V at a source voltage of $V_{DS} = 0.1$ V. Filled symbols show on-state, open symbols visualize off-state transistor.

Our preliminary simulation studies reported in (22) indicate that the effectiveness for S/D leakage suppression of the FG mainly depends on the effective electric permittivity of the material stack between the formed BG channel and the FG electrode. An increase of the body layer height under the FG electrode from 10 to 30 nm significantly degrades the simulated S/D leakage current suppression from ~8 aA/µm to ~2 nA/µm reducing the on-to-off current ratio for p- and n-type operation from 12 down to 3 decades. For this reason, a homogeneous recess fabrication with a defined body layer height in the FG region is a key process step in the fabrication of the device and has to be optimized in future work. As stated before, the devices discussed in this paper feature a relatively thick body layer of approximately 20 nm and devices with thinner body layer are currently fabricated. Also, scaling of the BG transistor length in future work is expected to result in further improvement of transistor performance, i.e. increased transconductance and drive current.

As already known from other MOSFET structures, a reduction of the front-gate length (FGL) considerably degrades the subthreshold slope and increases the threshold voltage for constant EOT values. Using a high-k oxide like hafnium oxide can mitigate this degradation as estimated in (22) by simulation.

Special to our device is the absence of a direct interface between the FG oxide and the channel. Additionally, no high temperature processing steps are required after FG oxide formation. Therefore, there are no technological handicaps like interface deterioration and temperature instabilities of high-k insulators which are known to degrade the device performance when high-k gate oxides like gadolinium oxide or other lanthanide oxides are introduced, as reported in references (10) and (11) for example. The fabrication of a high-k device is part of future work.

Effect of High Temperature on Transistor Performance

The fabricated device is electrically characterized at various temperatures ranging from 25 to 200°C. The measured input characteristics are depicted in Fig. 8 and Fig. 9. These measurements show an increasing on-current for the n-type mode FET and a decreasing on-current for the p-type FET with rising temperature (Fig. 9). The decrease can be assigned to a reduction of charge carrier mobility due to increasing phonon scattering within the silicon crystal lattice. The increase of on-current for the n-type mode

FET hints an increase of thermionic (field-) emission through the Schottky barriers as tunneling emission is known to be independent of temperature (12). This behavior can also be found in the measured BG transistor input characteristic under elevated temperatures for positive BG voltages in Fig. 10. As already stated before, this indicates that the mid-gap Schottky contacts are not perfectly symmetrical with respect to barrier heights for electrons and holes. By using Schottky barrier height engineering it is thought to be possible to compensate the temperature dependent decrease of carrier mobility by the increase of thermionic (field-) emission in order to realize an intrinsically temperature compensated transistor.

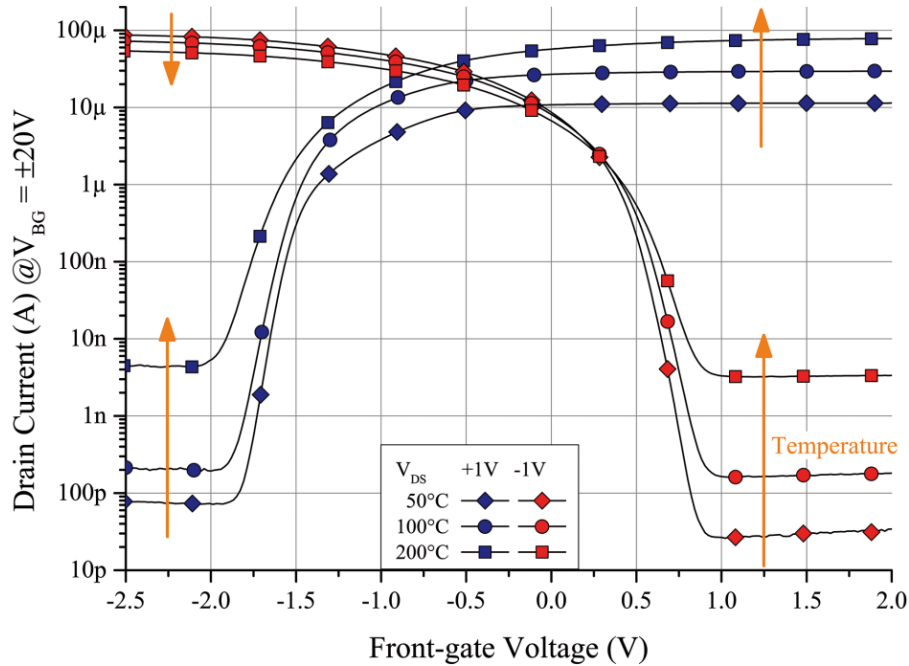

Fig. 8   Measured temperature effect on input characteristic of the fabricated front-gate transistor with an aluminum metal front-gate shown on a logarithmic scale. Transistor dimensions: BG length 33.9 µm, BG width 45.1 µm and FG length 18.9 µm.

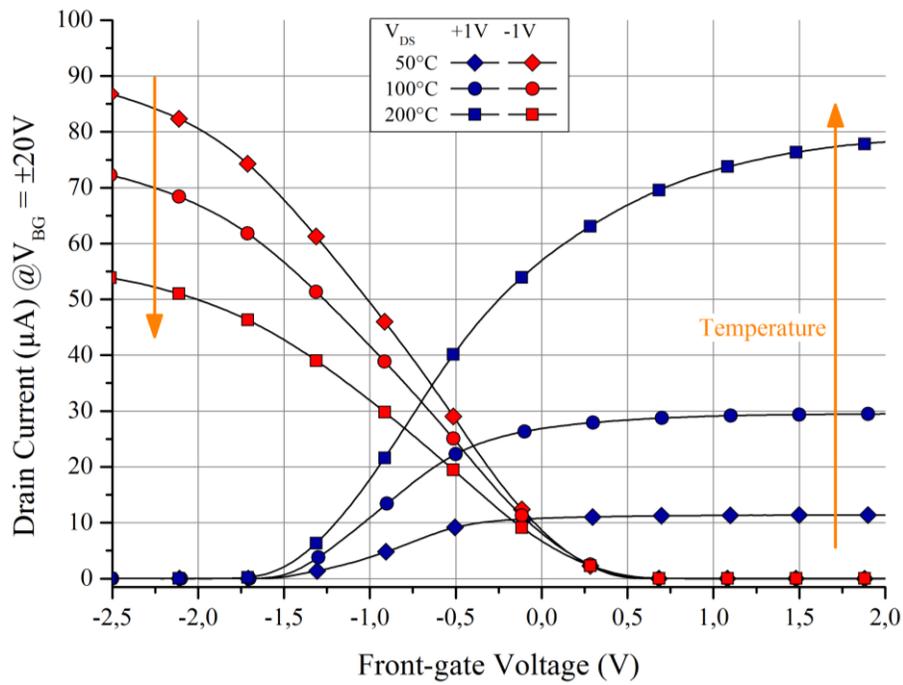

Fig. 9 Measured temperature effect on input characteristic of the fabricated front-gate transistor with an aluminum metal front-gate shown on a linear scale. Transistor dimensions: BG length 33.9 μm, BG width 45.1 μm and FG length 18.9 μm.

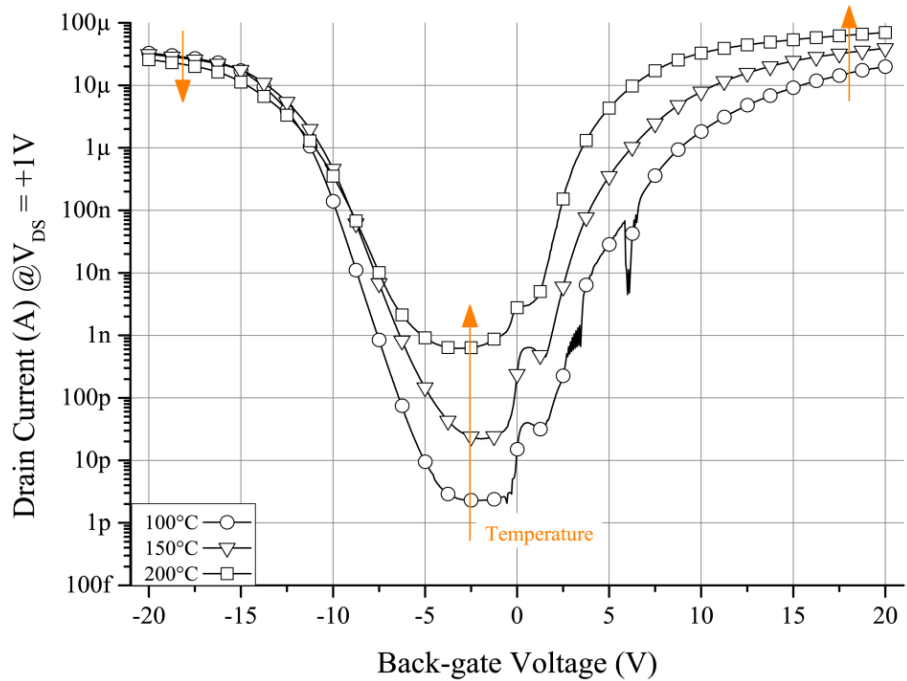

Fig. 10 Measured temperature effect on input characteristic of the fabricated back-gate transistor with shown on a logarithmic scale. Transistor dimensions: BG length 33.9 μm, BG width 45.1 μm.

A comparison of S/D leakage currents for different FET devices normalized to A/μm as a function of temperature is illustrated in Fig. 11. If available, long channel device data has been included. The off-state leakage current of the dehancement mode FET is significantly lower compared to reported SOIFET (23), JLFET (18) and FINFET (24) transistors. Furthermore, the simulation results of a DeFET device with 500/250 nm BGL/FGL and 5 nm body thickness are included to demonstrate the potential of future DeFET device generations. Note again, the device reported in this paper is limited by the relatively thick body layer of approximately 20 nm below the FG electrode.

The separation of carrier injection and channel formation from current control significantly suppresses parasitic S/D junction leakage and increases electrostatic control, particular at elevated temperatures. The potential barrier of the FG is sufficiently high to limit the off-current raise to less than 2 decades from 50°C to 200°C while keeping the on-to-off drain current ratio higher than 4 decades at 200°C (Fig. 8). The performance of the 1$^{st}$ generation planar DeFET devices is in the same order of magnitude as the previously fabricated Si-NW FET (4) clearly demonstrating the potential of the planar high temperature DeFET concept. Future devices will be based on ultrathin body SOI using body layer thicknesses of 5 to 10 nm below the FG electrode resulting in an exponential reduction of off-current drain leakage.

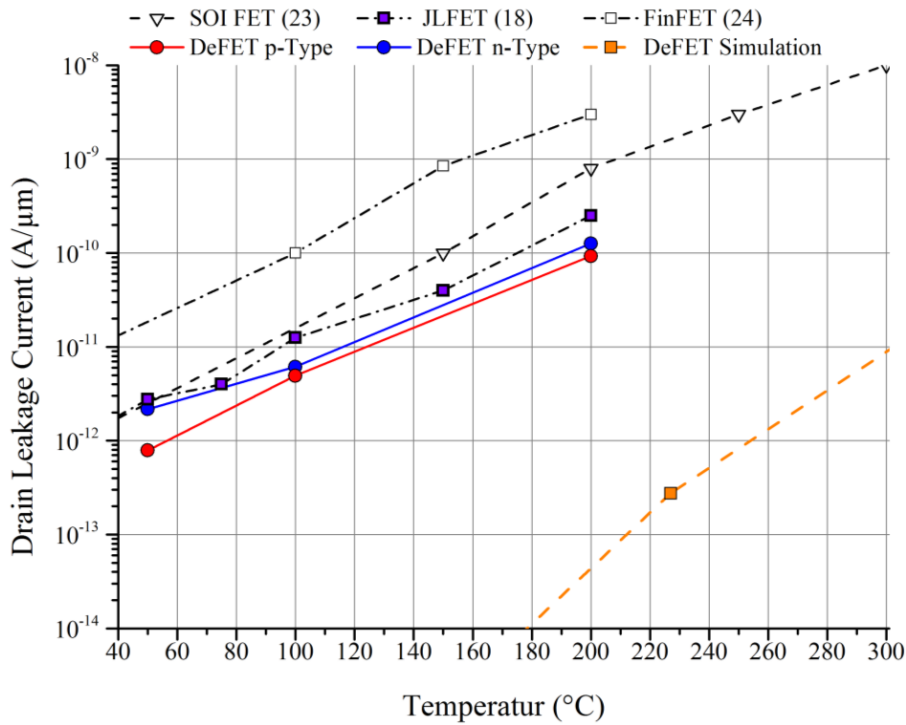

Fig. 11 Comparison of the temperature dependent S/D leakage current suppression of reported SOI MOSFET (23), JLFET (19), FinFET (24) and experimental as well as simulated dehancement mode FET.

## Conclusion

As a proof of concept, we fabricated a first version of a dopant-free long channel planar dehancement mode FET device. We found that the combination of an ambipolar SBFET for electrostatic doping and a unipolar JLFET for actual current flow control results in a superior on-to-off current ratio and leakage suppression at high temperatures. The device shows excellent on-to-off current ratio at 200°C of 4 decades combined with low leakage currents of 0.1 nA/µm. Based on the presented results, a device with intrinsic temperature compensation via balancing the positive temperature coefficient of the Schottky barriers against the negative temperature coefficient of the charge carrier mobility seems feasible. Furthermore, since no conventional doping is required, the device does not suffer from dopant dependent reduction of carrier mobility and parameter fluctuations as well. In addition, the degree of freedom to select instantly n- and p-type behavior via an ordinary electrical signal of the appropriate polarity on the BG allows designing reconfigurable circuits with increased functionality (5).

**Acknowledgment**

The authors would like to thank G. Hess, D. Noll and R. Heller for their assistance in device fabrication.